\documentclass[a4paper,11pt]{article}
\usepackage{longtable}
\usepackage[fleqn]{amsmath}
\usepackage{amssymb,amsthm}
\usepackage{graphicx}
\usepackage{bm}
\usepackage{footmisc}
\usepackage{afterpage}
\usepackage{float}
\usepackage{lscape}
\usepackage{longtable}

\usepackage[square,comma,sort&compress,numbers]{natbib}

\usepackage[dvipsnames]{xcolor}
\definecolor{gray75}{gray}{0.4}

\usepackage{fullpage}
\usepackage{etex}
\usepackage[utf8]{inputenc}
\usepackage[T1]{fontenc}
\usepackage{lmodern}
\usepackage[nolist,nohyperlinks]{acronym}
\usepackage{booktabs}
\usepackage{hepunits}
\usepackage{units}

\usepackage{multirow}

\usepackage{epigraph}
\epigraphsize{\small\itshape}
\usepackage{csquotes}

\usepackage{slashed}
\usepackage[section]{easy-todo}
\usepackage{caption}
\usepackage{subcaption}
\usepackage{setspace}
\usepackage{notoccite}
\usepackage{listings}
\usepackage{enumerate}

\usepackage[hyperfootnotes=false,pdfpagelabels]{hyperref}  

\hypersetup{%
    colorlinks=true, linktocpage=true, pdfstartpage=3, pdfstartview=FitV,%
    breaklinks=true, pdfpagemode=UseNone, pageanchor=true, pdfpagemode=UseOutlines,%
    plainpages=false, bookmarksnumbered, bookmarksopen=true, bookmarksopenlevel=1,%
    hypertexnames=true, pdfhighlight=/O,
   urlcolor=gray75, linkcolor=gray75, citecolor=gray75,
     pdftitle={Non-perturbative phenomenology},%
     pdfauthor={Leonardo Antunes Pedro},%
}

\usepackage{mciteplus}

\newcommand*{\bea}{\begin{eqnarray}}
\newcommand*{\eea}{\end{eqnarray}}
\newcommand*{\be}{\begin{equation}}
\newcommand*{\ee}{\end{equation}}

\newcommand*{\pref}[1]{(\ref{#1})}

\newcommand*{\nn}{\nonumber}

\newcommand*{\tr}{\mathrm{tr}}
\newcommand*{\diag}{\mathrm{diag}}

\newcommand{\bma}{\begin{pmatrix}}
\newcommand{\ema}{\end{pmatrix}}

\graphicspath{{./Images/}}

\renewcommand{\textflush}{flushepinormal}

\theoremstyle{plain}

\newtheorem*{rmk*}{Note}

\theoremstyle{definition}

\newtheorem*{defn*}{Definition}

\theoremstyle{remark}

\makeatletter
\renewcommand{\@epitext}[1]{
\itshape \begin{minipage}{\epigraphwidth}\begin{\textflush} #1
\end{\textflush}\end{minipage}\vspace{1ex}}

\makeatother
\title{Gauge invariance and the physical spectrum in the two-Higgs-doublet model}

\author{Axel Maas$^{1}$, Leonardo Pedro$^{2}$\\
$^1$Institute of Physics, NAWI Graz, University of Graz,\\
Universit\"atsplatz 5, A-8010 Graz, Austria\\
$^2$Centro de F\'isica Te\'orica de Part\'iculas, Universidade de Lisboa,\\
Av. Rovisco Pais, P-1049-001 Lisboa, Portugal}

\date{\today}

\begin{document}

\maketitle

\begin{abstract}
Observable states are gauge-invariant. In a non-Abelian gauge theory, these are necessarily composite operators. We investigate the spectrum of these operators in the two-Higgs-doublet model.
For this purpose, we are working along the lines of the Fr\"ohlich-Morchio-Strocchi mechanism to relate the physical spectrum to the spectrum of the elementary particles. We also investigate the consequences of spontaneous breaking of the global (custodial) symmetry group. Finally, we briefly comment on how to test the results using lattice methods.
\end{abstract}




\section{Introduction}

One requirement of particle physics theories is that their experimentally observable consequences must be gauge-invariant.
In Abelian gauge theories, this is achieved by a suitable dressing of elementary operators yielding gauge-invariant states and a gauge-invariant electric charge~\cite{Haag:1992hx,nonperturbativefoundations}. This additional dressing has a minor quantitative influence, and therefore a perturbative description using the elementary, gauge-dependent, electron and photon fields is successful. Nonetheless, objects like the hydrogen atom can still not be described perturbatively.

In non-Abelian gauge theories, the situation is more involved. Every gauge-invariant operator is necessarily composite, and gauge charges cannot be made gauge-invariant~\cite{Haag:1992hx}. For QCD, due to confinement, this does not surface as an additional complication, as only composite states, hadrons, can be observed anyway. The situation in the weak case is more subtle, due to the Brout-Englert-Higgs (BEH) effect.

Since the gauge symmetry remains---even in presence of the BEH effect---unbroken~\cite{Elitzur:1975im}, gauge-invariant states are still necessarily composite. These should be considered to be the relevant degrees of freedom~\cite{Banks:1979fi,'tHooft:1979bj}. This is emphasized by the fact that the (lattice-regularized) weak-Higgs sector shows no phase transition when moving into a region with QCD-like physics, i.\ e.\ exhibiting confinement of weak charges in the same way as in QCD~\cite{Osterwalder:1977pc,Fradkin:1978dv,Caudy:2007sf,Seiler:2015rwa}.

Somewhat surprisingly, the spectrum of the weak-Higgs sector is nonetheless described exceedingly well by the spectrum of the elementary, gauge-dependent Higgs and weak gauge boson fields in perturbation theory~\cite{Bohm:2001yx}. The explanation for this rests in a combination of the special structure of the Higgs sector together with the values of the parameters in the standard model, the Fr\"ohlich-Morchio-Strocchi (FMS) mechanism~\cite{Frohlich:1980gj,Frohlich:1981yi}. In particular in the Standard Model, there is no spontaneous symmetry breaking neither of the gauge nor of the global (custodial) symmetry group of the Higgs potential. It is therefore a-priori not clear whether in theories with a different Higgs sector a similar argument could be made, and therefore whether perturbation theory would be at all able to predict correctly the observable particle spectrum~\cite{Maas:2015gma,Torek:2015ssa}.

Arguably the simplest extension of the standard model which alters these structural properties are two-Higgs-doublet models (2HDM)~\cite{Branco:2011iw}, keeping the gauge group but changing the global symmetry group of the Higgs potential by adding a second Higgs doublet. Following the standard perturbative treatment\footnote{We follow the conventions used in the reference~\cite{signs} for the signs and constants.} of the two-Higgs-doublet model~\cite{accidental,*accidental2,Branco:2011iw,O'Neil:2009nr}, we will apply the FMS mechanism to this model and discuss lattice simulations to study it, in particular we will study its spectrum and the spontaneous symmetry breaking of the global symmetry group (which is indeed possible to occur in 2HDMs ~\cite{Lewis:2010ps}).

In section \ref{sec:operators} we will study the gauge-invariant operators in the 2HDMs. This is important not only to apply the FMS mechanism but also to prepare for lattice simulations. To handle various global symmetry groups of the Higgs potential, we employ the language of Majorana matrices and spinors~\cite{accidental,*accidental2}, reviewed in section~\ref{sec:majorana}.
After this, we will review and apply the FMS mechanism to the 2HDMs in section~\ref{sec:asymptotic}. We formulate spontaneous symmetry breaking in section \ref{sec:ssb}. We elaborate on the FMS mechanism in 2HDM in section \ref{sec:ssb}. Especially, we study the $Spin(4)$ symmetric potential and discuss the situations when a continuous or discrete symmetry group is spontaneously broken in section \ref{sec:potential}. We outline how the results and assumptions could be tested in lattice simulations of 2HDMs in section~\ref{sec:lattice}, which is possible in principle~\cite{Lewis:2010ps,Maas:2014nya}. This would complement other investigations of beyond-the-standard-model  physics (BSM) using lattice techniques \cite{DeGrand:2015zxa}. The extension to include photons and fermions will be discussed in sections \ref{sec:photons} and \ref{sec:fermions}, respectively. We conclude in section~\ref{sec:conclusions}.

\section{Gauge-invariant operators}
\label{sec:operators}

We want to know what are the  gauge invariant operators  of the 2HDM and among these the observable states that can be identified with the elementary gauge-dependent fields in perturbation theory (when the gauge is fixed).

Consider for the moment just the weak-Higgs sector: there are two weak $SU(2)_L$ Higgs doublets $\phi_1,\phi_2$ and the gauge field $W_\mu^j$ with $j,k,l=1,2,3$. The Lagrangian is: 
\begin{align*}
\mathcal{L}&\equiv ((D^\mu\phi_1)^\dagger(D_\mu\phi_1)+((D^\mu\phi_2)^\dagger(D_\mu\phi_2)-V(\phi_1,\phi_2)-\frac{1}{4}W_{\mu\nu}^jW^{j\mu\nu}\\
D_\mu&\equiv\partial_\mu+igW_\mu^j\frac{\sigma^j}{2}\\
W_{\mu\nu}^j&\equiv -\frac{i}{g}\tr([D_\mu,D_\nu]\sigma^j)=\partial_\mu W_\nu^j-\partial_\nu W_\mu^j-g\epsilon^{jkl}W^k_\mu W^l_\nu,
\end{align*}
where $V(\phi_1,\phi_2)$ is the Higgs Potential, $D_\mu$ is the covariant derivative dependent on the gauge field $W_\mu^a$, $W_{\mu\nu}^j$ is the gauge field strength tensor, $g$ is the coupling constant, $\epsilon^{jkl}$ is the Levi-Civita tensor, and $\sigma^j$ are the Pauli matrices in gauge space.

The construction is  more involved than for one Higgs doublet, as it is possible to construct more gauge-invariant composite operators from the two Higgs fields. 
In this paper we consider only polynomial operators\footnote{The result above can be extended \cite{invariantfunctions}: a generic function $f$ of the fields $\phi_1(x),\phi_2(x)$ and the Wilson lines $U(x,y,C)$ is a function of the  primitive operators listed in the basis of $16$ types, with no restriction on $f$ to be a polynomial. What this generically means can be found in \cite{invariantfunctions}. E.\ g.\ $f$ must have a finite number of arguments. For instance, the length of the Higgs field $\sqrt{\phi^\dagger_1\phi_1}$ is a function of the operator $\phi^\dagger_1\phi_1$, included in the above list.}. 
Any polynomial combination of $\phi_1,\phi_2$ which is gauge invariant is a polynomial on the 
linear independent inner products and skew-symmetric terms \cite{isotropy1,*isotropy2,*isotropy3}.
The inner products are $\phi_{1 a}^{*}\phi_1^{\ a}$, $\phi_{2a}^*\phi_2^{\ a}$, 
$\phi_{2a}^*\phi_1^{\ a}$; there are also the skew-symmetric combinations $\epsilon_{ab}\phi_1^{\ a} \phi_2^{b}$ and $\epsilon_{ab}\phi_1^{\ a *} \phi_2^{b *}$, with $\epsilon_{ab}$ the Levi-Civita symbol in 2 dimensions and $\phi_{j b}^{*}\equiv (\phi_{j}^{\ b})^{*}$ is the complex conjugate with $j=1,2$. 

Besides these composite operators involving fields at the same space-time points, it is possible to construct composite operators like $\phi^\dagger_1(x)U(x,y,C)\phi_1(y)$, where $U(x,y,C)$ is the parallel transport  from $y$ to $x$  along the path $C$. For an infinitesimal path the parallel transport involves the covariant derivative and for a point-like path this reduces to the previous set of operators. 

It is, of course, possible to construct gauge-invariant composite operators just from gauge fields, essentially $W$/$Z$-balls. 
Since these do not involve Higgs fields, they are the same as in Yang-Mills theory and therefore play no role here\footnote{The gauge-invariant content of the gauge field can be written as a function of the traces of Wilson loops, i.e. $tr(U(x,x,C'))$ with $C'$ a closed path~\cite{wilsonloops,*wilsonlines} (in classical field theory it is proved, in quantum field theory it is believed). Non-perturbatively, these operators also carry all the information about the bound-state spectrum, if they have non-zero overlap with all states~\cite{Wurtz:2013ova,Maas:2014pba}.}.

The basis of $16$ types of primitive (i.e. algebraically independent) gauge-invariant operators involving the Higgs fields is therefore:
\begin{itemize}
\item $\phi_{j}^{\dagger}(x)U(x,y,C)\phi_k(y)$
\item $\phi^\dagger_{j}(x)U(x,y,C)\overline{\phi}_k(y)$
\item $\overline{\phi}^\dagger_j(x) U(x,y,C)\phi_k(y)$
\item $\overline{\phi}^\dagger_j(x) U(x,y,C)\overline{\phi}_k(y)$
\end{itemize}
\noindent where  $\overline{\phi}^{\ a}_j(x)\equiv \epsilon^{ab}\phi_{jb}^*(x)$, the indices  $j,k=1,2$ are Higgs flavor indices and  $a,b=1,2$ are gauge indices.
The parallel transport $U(x,y,C)$ is from $y$ to $x$ following the path (line) $C$. It is an $SU(2)_L$ matrix (with no Higgs flavor indices) and so it satisfies
\begin{align*}
&U^a_{\ d}(x,y,C)=-\epsilon^{ab}U_b^{*c}(x,y,C)\epsilon_{cd},\ \mathrm{where}\  U_b^{*c}(x,y,C)\equiv (U^b_{\ c}(x,y,C))^*\\
&\overline{\phi}_{j}^\dagger(x) U(x,y,C)\overline{\phi}_{k}(y)=(\phi_{j}^\dagger(x) U(x,y,C)\phi_{k}(y))^*\\
&\overline{\phi}_{j}^\dagger(x) U(x,y,C)\phi_{k}(y)=-(\phi_{j}^\dagger(x) U(x,y,C)\overline{\phi}_{k}(y))^*.
\end{align*}
Therefore the above list includes the complex conjugates of all the operators of the list.


Note that for infinitesimal line elements
\begin{equation}
U(x,y,C)\approx (1+D_\mu(x)dl^\mu_1)(1+D_\nu(x)dl^\nu_2)...(1+D_\alpha(x)dl^\alpha_n)\nonumber
\end{equation}
\noindent where $dl_1,dl_2,...,dl_n$ (with $n$ finite) are infinitesimal Lorentz vectors which form the infinitesimal path $C$ by concatenation. In the following, we mainly consider the terms of order 1, yielding a scalar part, and of order $dl_1$, yielding a Lorentz vector part.

\section{Majorana construction}\label{sec:majorana}

In this section we write the possible gauge invariant operators in the language of Majorana spinors (representations of the symmetry of the Higgs doublets, not of the Lorentz group)~\cite{accidental}. We use matrices with well defined commutation relations instead of the Higgs flavor indices. 
It is useful for that purpose to review some consequences of generalizations of Pauli's theorem \cite{paulitheorem,*paulitheorem2}: 

Let $A^a$, $B^a$, $a\in\{1,...,2n\}$ with $n<4$ a natural number, be two sets of
$2^n\times 2^n$ complex unitary matrices satisfying
\begin{align*}
A^a A^b+A^b A^a&=2g^{ab}1\\
B^a B^b+B^b B^a&=2g^{ab}1
\end{align*}
where $g\equiv \diag(-1,...,+1,...)$, with $n$ entries equal to $-1$ and $n$ entries equal to $+1$. Then:
\begin{enumerate}\itemsep1pt \parskip0pt \parsep0pt
\item there is a complex unitary matrix $S$ such that
$B^a=S A^a S^{-1}$, for all $a\in\{1,...,2n\}$. 
$S$ is unique up to a phase;
\item  there is a basis where all $A^a$ are real.
If $A^a$ and $B^a$ are all real, then $S$ can be made real;
\item  the Clifford algebra over the complex (resp. real) numbers 
generated by the matrices $A^a$ is isomorphic to the algebra of the 
$2^n\times 2^n$ complex (resp. real) matrices.
\end{enumerate}

\noindent Also useful is the conjugation operator $\Theta$, an anti-linear involution commuting with the matrices $A^a$. It follows from the above theorem that $\Theta$ is unique up to a complex phase. The set of Majorana spinors is the set of $2^n$ dimensional complex vectors $u$ satisfying the Majorana condition (defined up to a complex phase) $\Theta u= u$. The set of Majorana spinors is then a $2^n$ dimensional real vector space. Note that linear combinations of Majorana spinors with complex prefactors  in general do not  not satisfy the Majorana condition.

In the following we will consider different dimensions $2^n$ for the Majorana spinors.
For $n=1$, we define the Pauli matrices as $\sigma^1\equiv A^2$, $\sigma^2\equiv iA^1$, $\sigma^3\equiv -i\sigma^1\sigma^2=A^2A^1$. The Pauli spinor is a 2-dimensional complex vector.

For $n=2$  let $\Phi\equiv\frac{1}{\sqrt{2}} \left[ \begin{smallmatrix} \overline{\xi}\\ \xi \end{smallmatrix} \right]$ be a Majorana spinor satisfying the Majorana condition $(i\sigma_2\otimes i\sigma_2)\Phi^*=\Phi$, implying $\overline{\xi}=i\sigma_2\xi^*$ where $\xi$ is a Pauli spinor.
The Pauli matrix on the right acts on the Pauli spinors $\xi$, $\overline{\xi}$, the one on the left acts on the space $\left[ \begin{smallmatrix}\overline{\xi}\\\xi \end{smallmatrix} \right]$.  Below it is illustrated how the usual Higgs doublet is rewritten as a Majorana spinor, and how both gauge and custodial transformations act\footnote{For $n=2$ we identify $A^1=1\otimes i\sigma_1$, $A^2=1\otimes i\sigma_2$, $A^3A^4=i\sigma_3 \otimes 1$, $A^4A^5=i\sigma_1 \otimes 1$.
The $SU(2)_L$ group generators are $A^a$ while the matrices $[A^a,A^b]$ ($a,b=3,4,5$) form a basis for the skew-hermitian matrices invariant under $SU(2)_L$. These matrices form a $Spin(3)$  group, i.e. the $SU(2)_R$ custodial group~\cite{Branco:2011iw}.

Take a Pauli spinor $\xi=\left[ \begin{smallmatrix}a+ib\\c+id \end{smallmatrix} \right]$, with $a,b,c,d$ real numbers. Then it is mapped on a Majorana spinor\\ 
$\sqrt{2}\Phi=\left[ \begin{smallmatrix} \overline{\xi}\\ \xi \end{smallmatrix} \right]=
a e_1
+b e_2
+c e_3+
d e_4$,
where the basis vectors are the columns of the matrix:
\begin{align*}
U^\dagger\equiv \frac{1}{\sqrt{2}}\left[ \begin{smallmatrix} e_1 & e_2 & e_3 & e_4 \end{smallmatrix}\right]=
\left[ \begin{smallmatrix}
0 & 0 & 1 & -i\\
-1 & i & 0 0\\
1 & i & 0 & 0\\
0 & 0 & 1 & i\end{smallmatrix} \right],
\mathrm{which\ changes\ to\ a\ real\ basis\ } 
U\Phi=\left[ \begin{smallmatrix} a\\b\\c\\d \end{smallmatrix} \right]
\end{align*}
\\
We have 
$(i\sigma_2\otimes i\sigma_2)=U^\dagger U^*$ 
and so the Majorana condition is the real condition $U^*\Phi^*=U\Phi$.\\
The Majorana condition is equivalently $\Phi^{bc}=\epsilon^{bd}\epsilon^{cf}\Phi_{df}^{*}$ where $\Phi_{bc}^{*}\equiv (\Phi^{bc})^{*}$.}.

For the 2HDM, there is also a flavor space and thus an 8-dimensional Majorana spinor is necessary. Consider for $n=3$,  $\phi\equiv \left[ \begin{smallmatrix} \Phi_1\\ \Phi_2\end{smallmatrix} \right]$, where $\Phi_{1,2}$ are the previous 4-dimensional Majorana spinors. Then $\phi$ satisfies the Majorana condition $(1\otimes i\sigma_2\otimes i\sigma_2)\phi^*=\phi$.  In this expression, the Pauli matrix on the right acts on the Pauli spinors, the one in the middle acts on the custodial space $\left[ \begin{smallmatrix}\widetilde{\xi}\\ \xi \end{smallmatrix} \right]$, while the one on the left acts on the flavor space $\left[ \begin{smallmatrix}\Phi_1 \\\Phi_2 \end{smallmatrix} \right]$.

We define now $\Sigma_j\equiv A^{j+3}$ ($j=1,2,3$), $\Sigma_4\equiv A^1 A^{2}A^{3}$ and  $\Sigma_5\equiv  \Sigma_1 \Sigma_2\Sigma_3\Sigma_4=-A^7$. There is a basis where $\epsilon_{jkl} A_{k}A_{l}= 1\otimes 1\otimes i\sigma_j$, $\epsilon_{jkl} \Sigma_{k}\Sigma_{l}= 1 \otimes i\sigma_j \otimes 1$, $\Sigma_{5}=\sigma_3 \otimes 1 \otimes 1$ and $\Sigma_{4}= \sigma_1 \otimes 1 \otimes 1$.

The generators of the gauge transformations $SU(2)_L$ are $\epsilon_{jkl} A_{k}A_{l}= 1\otimes 1\otimes i\sigma_j$. We use the shorter notation $i\sigma_j\phi\equiv(1\otimes 1\otimes i\sigma_j)\phi$.

The matrices $1, \Sigma_a$ $a=1,...,5$ form a basis for the hermitian matrices conserved by the generators of $SU(2)_L$. Note that the $\Sigma_a$ anti-commute with each other. The matrices $[\Sigma_a,\Sigma_b]$ form a basis for the skew-hermitian matrices conserved by the generators of $SU(2)_L$ and are the generators of a $Spin(5)$ group\footnote{the full group is $(Spin(5)\times SU(2)_L)/Z_2$;
$Spin(5)$ is the double cover of $SO(5)$; $[\Sigma_a,\Sigma_b]\equiv\Sigma_a\Sigma_b-\Sigma_b\Sigma_a$ is the commutator;\\
$1=\frac{1}{5!}\epsilon_{abcdf}\Sigma_a\Sigma_b\Sigma_c\Sigma_d\Sigma_{f}$ and so $\Sigma_a=\frac{1}{4!}\epsilon_{abcdf}\Sigma_b\Sigma_c\Sigma_d\Sigma_{f}$;\\
$U(x,y,C)$ acts on the gauge indices and so it commutes with $\Sigma_a$}.

Therefore we rewrite the basis of $16$ types of primitive composite operators as
\begin{itemize}
\item $\phi^{\dagger}(x)U(x,y,C)\phi(y)$ (singlet under $SO(5)$);
\item $\phi^{\dagger}(x)U(x,y,C)\Sigma_a\phi(y)$ ($\mathbf{5}$ representation of $SO(5)$);
\item $\phi^{\dagger}(x)U(x,y,C)[\Sigma_a,\Sigma_b]\phi(y),\ a,b=1,...,5$ ($\mathbf{10}$ representation of $SO(5)$);
\end{itemize}

\section{Observable states of the two-Higgs-doublet model}\label{sec:asymptotic}

We can now evaluate the spectrum of the 2HDMs.  For this purpose, we specify the Higgs potential of the model and its symmetry group, following the basis-invariant formalism~\cite{O'Neil:2009nr}.
\begin{equation*}
V(\phi)=\mu_a\phi^\dagger\Sigma_a\phi+\frac{1}{2}\lambda_{ab}(\phi^\dagger\Sigma_a\phi)(\phi^\dagger\Sigma_b\phi)
\end{equation*}
where $a,b=0,1,...5$ and $\Sigma_0\equiv 1$. 

If we promote the parameters of the Higgs potential to 
background fields\footnote{A (non-dynamical) background field or spurion enters in the definition of the Lagrangian but it is not a variable of the Lagrangian.  When calculating the observables, the background fields are replaced by numerical values. It is a representation of a group of background symmetries of the Lagrangian, but there are no Noether's currents associated with such background symmetries if the numerical values are non-null. The observables are invariant under the action of the group of the background symmetries. See \cite{group2HDM2,*group2HDM,*Botella:2012ab,*georgi} for more details and related approaches.}, the Lagrangian is invariant under the gauge group $SU(2)_L$ and the group of background symmetries $Spin(5)$ with generators $[\Sigma_a,\Sigma_b]$. The gauge-invariant composite operators are therefore classified according to the representations of the global $Spin(5)$ group. The elementary $8$-dimensional spinor $\phi$ containing the Higgs fields is the tensor product of a $4$-dimensional complex representation of $Spin(5)$ and a $2$-dimensional complex representation of $SU(2)_L$, verifying a Majorana condition. Hence for $a,b\neq 0$, $\mu_0,\lambda_{00}$ are singlets, $\mu_a,\lambda_{0a}$ are $5$-dimensional representations of $SO(5)$ and $\lambda_{ab}$ is a tensor of $SO(5)$. 

Let now $V(\phi=\frac{v}{\sqrt{2}}\phi_0)$ be an absolute minimum of the classical potential where $v$ is the vacuum expectation value (vev) and $\phi_0^\dagger\phi_0=1$. Without loss of generality due to the background symmetry, by reparametrization of the Higgs potential we assume that $\Sigma_5\phi_0=\phi_0$. Such a condition involving $\Sigma_5$ is not invariant for the generators $[\Sigma_5,\Sigma_a]$ and so it breaks the background symmetry $Spin(5)\to Spin(4)$. The consequences of this breaking will be discussed in the next section.
We define $H_1\equiv\frac{1+\Sigma_5}{2}\phi$ and $H_2\equiv\Sigma_4\frac{1-\Sigma_5}{2}\phi$, hence if $\phi=\frac{v}{\sqrt{2}}\phi_0$ then $H_2=0$. Note that $H_1,H_2$ are not complex doublets but $4$ dimensional Majorana spinors.
In table \ref{table:states} 16 primitive operators are listed. There the isomorphism $Spin(4)\simeq(SU(2)_{R1}\times SU(2)_{R2})$ with the $SU(2)_{R1}$ generators $\Sigma_j\Sigma_4(1+\Sigma_5)/2$ and the $SU(2)_{R2}$ generators $\Sigma_j\Sigma_4(1-\Sigma_5)/2$ was used to classify the states.

\begin{table}[htb!]
\begin{center}
\begin{tabular}{c c c c c c c}
Lorentz rep. & $J (SU(2)_{R1})$ & $J (SU(2)_{R2})$  & Operator & Expansion\\
\hline
scalar & 0 & 0 &  $H^\dagger_1 H_1$ &  $\frac{v^2}{2}+v\phi_{0}^\dagger\varphi$\\
scalar & 0 & 0 &  $H^\dagger_2 H_2$ & 0 \\
scalar & 1/2 & 1/2 &  $H^\dagger_1 \Sigma_a\Sigma_4 H_2$ &  $\frac{v}{2}\phi_{0}^\dagger\Sigma_a\varphi$\\
vector & 1 & 0 &  $H^\dagger_1 D_\mu \Sigma_{j}\Sigma_{4} H_1$ &  $\frac{gv^2}{4} W_\mu^j$\\
vector & 0 & 1 &  $H^\dagger_2 D_\mu \Sigma_{j}\Sigma_{4} H_2$ & 0 \\
vector & 1/2 & 1/2  &  $H^\dagger_1 D_\mu \Sigma_{a}\Sigma_5\Sigma_4 H_2+H^\dagger_2 D_\mu \Sigma_{a}\Sigma_5\Sigma_4 H_1$ & 0 
\end{tabular}
\caption{\label{table:states}Gauge-invariant states corresponding to the elementary states of 2HDM classified by the custodial symmetry $Spin(4)\simeq(SU(2)_{R1}\times SU(2)_{R2})$ \label{TAB:states}, where the potential has an absolute minimum for a minimum satisfying $\Sigma_5\phi=\phi$. The indices are $j=1,2,3$ and $a=1,2,3,4$. Note that for the expansion the gauge was fixed such that\\$\sqrt{2}\phi(x)=v \phi_0+\varphi(x)$;\hspace{.5cm} $\Sigma_5\phi_0=\phi_0$;\hspace{.5cm} $i\sigma_j\phi_0=\Sigma_4 \Sigma_j\phi_0$;\hspace{.5cm} $\phi_0^\dagger\phi_0=1$.
}
\end{center}
\end{table}

After a complete gauge fixing in a suitable gauge~\cite{Frohlich:1980gj,Frohlich:1981yi,Maas:2012ct} the BEH effect allow us to expand the Higgs field $\sqrt{2}\phi=v \phi_0+\varphi$ around a constant reference point $\frac{v}{\sqrt{2}} \phi_0$ minimizing the Higgs potential. We assume now that the fluctuations $\varphi$ around the vacuum are generically small compared to $v$. The reference point is chosen to obey:
\begin{align*}
i\sigma_j\phi_0=\Sigma_4 \Sigma_j\phi_0\ (j=1,2,3)
\end{align*} 
Thus, $\phi^0$ conserves a $SO(3)\times Spin(3)\simeq(SU(2)_{R1}\times SU(2)_{R2})/Z_2$ symmetry, whose generators are $(\Sigma_4 \Sigma_j(1+\Sigma_5)/2-i\sigma_j)$ and $\Sigma_4 \Sigma_j(1-\Sigma_5)$, respectively.

We will use the reference point to fix a system of coordinates for the gauge-dependent elementary fields.

The 4 projections $\phi_0\phi^\dagger_0$ and $-\Sigma_4\Sigma_j\phi_0\phi_0^\dagger\Sigma_4\Sigma_j$ (for fixed $j=1,2,3$) which sum to the identity allow us to decompose the 4 dimensional real spinor representation space of $SU(2)_L$ into 4 real subspaces of dimension 1. 
In the subspace proportional to $\phi_0$, we have the fields $\phi^\dagger_0 H_1$, $\phi^\dagger_0 H_2$.
In the subspace proportional to $\Sigma_4\Sigma_j\phi_0$, we have the would-be Goldstone bosons
$\phi^\dagger_0\Sigma_4\Sigma_jH_1$ and, in addition, $\phi^\dagger_0\Sigma_4\Sigma_jH_2$.


For the triplet representations of $SU(2)_L$ we have the  projections 
$\frac{1}{4}(\phi_0\otimes \Sigma_4\Sigma_j\phi_0 -\Sigma_4\Sigma_j\phi_0 \otimes \phi_0)(\phi^\dagger_0\otimes \phi^\dagger_0 \Sigma_4\Sigma_j-\phi^\dagger_0\Sigma_4\Sigma_j \otimes \phi^\dagger_0)$
which decompose the 3 dimensional real representation space into 3 real subspaces of dimension 1.
In the subspace proportional to 
$(\phi_0\otimes \Sigma_4\Sigma_j\phi_0 -\Sigma_4\Sigma_j\phi_0 \otimes \phi_0)$ we have in this gauge the field $\phi^\dagger_0 D_\mu \Sigma_4\Sigma_j\phi_0=\frac{g}{2}W_\mu^j$.

Keeping only the first terms involving up to one elementary field in the expansion around the reference point
\begin{align*}
H^\dagger_1 H_1&\approx \frac{v^2}{2}+v\phi_0^\dagger\varphi\\
H^\dagger_1 H_2&\approx \frac{v}{2}\phi_0^\dagger\Sigma_4\varphi\\
H^\dagger_1\Sigma_j\Sigma_4 H_2 &\approx \frac{v}{2}\phi_0^\dagger\Sigma_j\varphi\\
H^\dagger_1 D_\mu \Sigma_{j}\Sigma_{4} H_1&\approx \frac{g v^2}{4} W_\mu^j\ (j=1,2,3),
\end{align*}
\noindent as a generalization of the 1HDM~\cite{Frohlich:1980gj,Frohlich:1981yi}. It is straightforward to see that all other possibilities from the basis of primitive invariants involving at most one covariant derivative expand to two or more elementary fields at leading order, 
since the vacuum expectation value satisfies $\Sigma_5\phi_0=\phi_0$ and so its contribution to $H_2$ is null.
It is possible to construct states with different Lorentz representation than those considered, using further covariant derivatives, but such states cannot expand to a single elementary field, as there are none with other Lorentz quantum numbers.


\section{Spontaneous symmetry breaking in 2HDMs}
\label{sec:ssb}

We saw in the previous section that by choosing a reference point minimizing the Higgs potential we necessarily break the background symmetry $Spin(5)\to Spin(4)$.
Therefore, if the absolute minimum is not unique (up to gauge transformations) such a choice is necessarily in conflict with a symmetry of the model.
There are then two possibilities: either the symmetry of the model is spontaneously broken or it is not. Once the model is chosen, whether there are spontaneously broken global symmetries 
or not is a dynamical phenomenon, requiring suitable calculational methods to test. It may indeed occur in 2HDMs~\cite{Lewis:2010ps} depending on the Higgs potential. 
Without further information we can only assume that it occurs or that it does not occur.

We assume from now on that  
whenever the choice of an absolute minimum is in conflict with a global symmetry of the model, such symmetry is spontaneously broken. Then the correspondence established in the previous section is valid, since in the picture where spontaneous symmetry breaking is a particular case of explicit symmetry breaking~\cite{Lewis:2010ps}
the conflict is avoided as such a would-be global symmetry of the model is explicitly broken by an infinitesimal parameter.

%
%

But such an assumption must be confirmed. In section \ref{s:spont} we study particular 2HDMs and evaluate the consequences for the spectrum for the possibility that the assumption is not valid.

Note that the definition of spontaneous symmetry breaking crucially depends on the physically realizable operations~\cite{symmetrybreaking,*wignerspontaneous,*Perez:2008fv}.

\section{The FMS mechanism}
\label{sec:fms}
The FMS mechanism establishes that there is a correspondence between the elementary gauge-dependent fields and the primitive composite states obtained by replacing the reference point $\frac{v}{\sqrt{2}}\phi_0$ (used to fix the gauge-dependent coordinate system) by the field $H_1$. The correspondence is one-to-one, except for the would-be Goldstone bosons $\phi^\dagger_0\Sigma_4\Sigma_jH_1$ which disappear from the spectrum, since $\Sigma_4\Sigma_j$ is skew-adjoint and therefore $H_1^\dagger \Sigma_4\Sigma_jH_1=0$. As we have seen in section~\ref{sec:asymptotic}, this correspondence applies to the 2HDMs---under the assumption of spontaneous symmetry breaking for non-unique minima of the potential.

Such correspondence becomes an equality if the field fluctuations become small enough, compared to the vev. Consider for instance the complete expansion of the scalar operator
\begin{equation*}
2H_1^\dagger H_1= v^2+2v\phi_0^\dagger\varphi+\varphi^\dagger\varphi.
\end{equation*}

A correlator of this gauge-invariant operator would yield to leading order just the propagator of the fluctuation field $\phi_0^\dagger\varphi$. Since the mass is given by the poles, to this order the composite state will have the same mass $m_H$ as the elementary state~\cite{Frohlich:1980gj,Frohlich:1981yi}.
This explains why the physical composite scalar operator has the same mass as the elementary Higgs field.
A similar argument can be made for the gauge bosons~\cite{Frohlich:1980gj,Frohlich:1981yi}. 
Therefore, the spectrum harbors a physical vector triplet with the same mass as the elementary gauge bosons. The  would-be Goldstone bosons $\phi_0^\dagger \Sigma_j \Sigma_4\varphi$ constitute the longitudinal degrees of freedom of $W_\mu^j$. 
For states with quantum numbers where there is no leading term corresponding to any elementary particle, the first contribution comes from scattering states.
 Of course, it is possible to doubt the correctness of the expansion\footnote{The operators $v\phi_0^\dagger\varphi$ and $\varphi^\dagger\varphi$ have the same quantum numbers hence they cannot be distinguished, except in an approximate way by the energy spectrum or in perturbation theory (in Quantum Electrodynamics we should also sum all possible initial and final states including those with soft photons in a finite energy window, to avoid infrared divergences \cite{kln1,*kln2,*kln}).
The operator $\varphi^\dagger\varphi$ is assumed to be a scattering state such that its energy spectrum starts at $\sim 2 m_H$, despite that it involves interactions:
\begin{itemize}
\footnotesize
\item the Higgs decay width is of the order of a few \MeV
\item the binding energy of the $SU(2)_L$ gauge interactions is expected to be below $0.1~\GeV$ if it exists at all (consider for instance a positronium where the electron mass is replaced by the Higgs mass and the coupling constant is replaced by the weak coupling constant).
\item the binding energy of the interactions from the Higgs potential is extremely weak
when it exists at all~\cite{selfcoupling,selfcoupling2,selfcoupling3} 
(for the parameters' scale around the Standard Model) 
and so the energy spectrum starts at $\sim 2 m_H$. Moreover no evidence that such bound states exist have yet been found for the Standard Model~\cite{Wurtz:2013ova,Maas:2014pba}.
\end{itemize}
In general, since the Higgs is among the most heavy gauge-dependent elementary fields, the energy spectrum of the next-to-leading contributions starts far from the mass of the gauge-dependent elementary field and so are negligible (w.r.t the leading contribution) for the state's mass near the mass of the elementary field.  In standard perturbation theory the  mass of the asymptotic states is the mass of the elementary fields, for the remaining intermediate states we do not expect deviations since the (gauge-invariant) Lagrangian is the same---unless there are new bound states or other unexpected non-perturbative effects.}. But the prediction has been confirmed non-perturbatively in various lattice calculations for the weak-Higgs theory with one doublet~\cite{Maas:2012tj,Maas:2013aia,Wurtz:2013ova,Maas:2014pba}.
Thus, the FMS mechanism appears to be indeed the correct description of the electroweak theory.

At loop level, where renormalization scheme issues affect the poles on the right-hand side, the situation becomes more involved, and it is not yet fully developed~\cite{Frohlich:1980gj,Frohlich:1981yi}.
It remains to be checked the contribution in perturbation theory from the next-to leading terms of the gauge-invariant states, since there are measured precision electroweak observables which must be accounted for. However, there are theoretical arguments indicating that the standard perturbative expansion assuming a gauge-dependent vacuum
expectation value cannot be asymptotic to gauge-dependent correlation functions~\cite{Frohlich:1981yi}. Thus, the standard perturbative expansion may still not fully capture all features, though the consequences of this are likely quantitatively irrelevant for the standard model.

The expansion of $H_1^\dagger \Sigma_a\Sigma_4 H_2$  $(a=1,...,4)$ selects the components of the second Higgs doublet, and thus the spectrum contains a quadruplet of particles with the same masses as the fields of the second Higgs doublet. Since no other operator has a non-vanishing leading order, this completes the spectrum. Thus, for the 2HDMs under the assumptions of no spontaneous global symmetry breaking and that the next-to-leading terms do not lead to significant deviations, the FMS mechanism predicts, as for the standard-model case, a coincidence of the perturbative and physical spectrum. Of course these assumptions and the FMS mechanism must be validated non-perturbatively in 2HDMs, a point we will return to in section \ref{sec:lattice}.

The FMS mechanism can be extended to fermions~\cite{Frohlich:1980gj,Frohlich:1981yi}, yielding
\begin{equation*}
\sqrt{2}H^\dagger_1\Psi= v\phi_0^\dagger\Psi+\varphi\Psi,
\end{equation*}
where $\phi_0^\dagger\Psi$ is a fermion field, e.g. an electron with left chirality. Thus, composite operators of fermions and a single Higgs particle yield a gauge-invariant description of the fermions in the standard model, with the same mass at leading order. This is possible due to the scalar nature of the Higgs, which does not alter the spin or parity of the states. However, due to the intrinsic problem with chiral gauge theories on the lattice, not to mention the computational costs for even a moderately extended mass hierarchy, there is not yet any numerical evidence for this correspondence in the full standard model, or even just a subset of the fermion sector. The extension of the FMS mechanism to include photons and fermions in the 2HDMs will be discussed in sections \ref{sec:photons} and \ref{sec:fermions}, respectively.

\section{The $Spin(4)$ symmetric 2HDM}\label{s:spont}\label{sec:potential}

Now, we generalize the statements of the previous section.

The most general $Spin(4)$ symmetric potential is
\begin{align}
V(\phi)=\mu_0\phi^\dagger \phi+\mu_5\phi^\dagger\Sigma_5\phi+\frac{1}{2}\lambda_{00}(\phi^\dagger\phi)^2+\lambda_{05}(\phi^\dagger\phi)(\phi^\dagger\Sigma_5\phi)+\frac{1}{2}\lambda_{55}(\phi^\dagger\Sigma_5\phi)^2\label{potential2}
\end{align}
An analysis of the minima structure of the above potential is given in appendix~\ref{a:minima}. To avoid breaking the $Spin(4)$ group spontaneously we assume that the minimum the potential satisfies $\pm \Sigma_5\phi=\phi$.

We will assume from now on that $\lambda_{05}=0$ and we will study three particular cases in detail: 
\begin{enumerate}
\item the Maximally-Symmetric where $\mu_{5}>0$ and $\lambda_{55}=0$ has a unique minimum (up to gauge transformations) and its phenomenology is well studied and viable (in the sense that it is not ruled out by experiments) \cite{MS2HDM}. It will be used as a kind of ``control sample'' since no surprises are expected from the lattice simulations in the parameter space allowed by the experiments in comparison with perturbation theory, as also discussed in the previous section.
\item the limit $\mu_{5}\to 0$ with $\mu_{5}>0$ and $\lambda_{55}\neq 0$ will be used for a study of
spontaneous symmetry breaking of the $Z_2$ discrete symmetry which appears when $\mu_5=0$ and $\lambda_{55}\neq 0$. 
Since discrete symmetries break without Goldstone bosons, the spectrum of this theory is expected to be similar as in the maximally symmetric case.
\item the limit $\mu_{5}\to 0$ with $\mu_{5}>0$  and $\lambda_{55}=0$ will be used to study
spontaneous symmetry breaking of the continuous $Spin(5)\to Spin(4)$ symmetry which appears when $\mu_5=0$ and $\lambda_{55}=0$. 
According to Goldstone's theorem we expect $4$ massless Goldstone bosons.
\end{enumerate}

\noindent Without loss of generality we assume $\mu_5\geq 0$ (we can change its sign by a background symmetry transformation $\phi\to \Sigma_4\Sigma_5\phi$).
We then have that $\lambda_{00}=(m_h^2+m_H^2-2\mu_5)/v^2$, $\lambda_{55}=(-m_H^2+2\mu_5)/v^2$, $\mu_0=\frac{m_h^2}{2}-\mu_5$ and for the absolute minimum of the potential $\Sigma_5\phi_0=\phi_0$.
The terms in $\mu_5,\lambda_{55}$ break (softly if $\lambda_{55}=0$) the symmetry $Spin(5)\to Spin(4)$~\cite{accidental}, giving the same mass $m_H$ to the Higgs states $\phi_0^\dagger\Sigma_a\varphi$ ($a=1,2,3,4$) which are now mass eigenstates~\cite{O'Neil:2009nr}---these states are related to the states $H^\pm$, $R$ and $I$ defined in section~\ref{sec:photons}. The mass of the Higgs boson $\phi_0^\dagger\varphi$ is $m_h$ while the mass of the $W$ gauge triplet is $m_W\equiv gv/2$ (at tree level).
When $\lambda_{55}=0$, then $4\mu_5=2m_H^2$ 
so $(m_H^2+m_h^2-\sqrt{m_H^4+m_h^4})<4\mu_5<(m_H^2+m_h^2+\sqrt{m_H^4+m_h^4})$ and there is only one minimum.
When $\mu_5=0$ then a discrete symmetry $Z_2$ appears, the group is $Spin(4)\rtimes Z_2$ 
with the $Z_2$ transformation $\phi\to \Sigma_4\phi$ and the subgroup $Spin(4)$ is a normal subgroup. 

\paragraph{Breaking of a continuous symmetry?}

A conceptual interesting question is what happens if a (global) continuous symmetry of the Higgs potential is spontaneously broken, which is indeed possible~\cite{Lewis:2010ps}. In a standard perturbative analysis, this will give rise to massless Goldstone bosons originating from the Higgs doublets. These would be part of the spectrum, and would therefore add additional light particles to the spectrum, which could be interpreted, e.\ g., as axions. It is therefore interesting to see how this translates in the FMS perspective.

Since the unbroken subgroup is the $Spin(4)$ group, the observable states are given in table~\ref{table:states}.
First consider the quadruplet
\begin{align*}
H_1^\dagger  \epsilon_{jkl}\Sigma_a\Sigma_4 H_2\nn,
\end{align*}
This operator selects the four components of $H_2$ which turn out in this basis to be just the additional Goldstone bosons, as can be read off from the potential. Hence, this gauge-invariant operator indeed carries the information on the physical Goldstones, which also have been observed in lattice calculations for a different symmetry~\cite{Lewis:2010ps}.

This leaves only the vector states. The operators are
\begin{align*}
&H_1^\dagger D_\mu [\Sigma_j,\Sigma_k] H_1\\
&H_2^\dagger D_\mu [\Sigma_j,\Sigma_k] H_2\\
&H_1^\dagger D_\mu [\Sigma_a,\Sigma_5] (\Sigma_4 H_2)
\end{align*}
The first two are each triplets under the $SU(2)_{R1,2}$ subgroups. The last one is a quadruplet. Since only $H_1$ expands to a non-zero value, only the first operator yields a triplet with the mass of the $W$ bosons, and all other vanishes. This is a particular nice manifestation of the symmetry breaking pattern, as the 10 operators in the multiplet of the broken symmetry are no longer degenerate, as only one yields massive states and the other two scattering states.

Thus, in the case of a spontaneous breakdown of the global symmetry group the physical spectrum coincides with the one in perturbation theory. However, if the assumption of spontaneous symmetry breaking is not valid then all correlators should give an identical result and so we will have a multiplication of  vector degrees of freedom 
(as was pointed out in~\cite{Maas:2015gma}, where the absence of spontaneous symmetry breaking was implicitly assumed).

\paragraph{Breaking of a discrete symmetry?}
Up until now we assumed that the potential has only one continuous connected set of minima, like in the one-Higgs-doublet case. In the standard model case, this is the only possibility. But in the 2HDM case, it is possible to have not continuously connected sets of absolute minima.

This situation is found for the potential \pref{potential2} for the case $\lambda_{05}=\mu_{5}=0$~\cite{accidental}. In this case actually an additional global $Z_2$ in the custodial symmetry arises, yielding a $Spin(4)\times Z_2$ symmetry group. There are then different symmetric sets of absolute minima, not continuously connected by gauge transformations. These sets of minima are related by the $Z_2$ subgroup, and cannot be continuously deformed into each other.

Their presence has great significance for the physical spectra in the absence of the spontaneous symmetry breaking. It essentially implies that the multiplets of the $SU(2)_{R1,R2}$ groups are symmetric under exchanges of the respective  groups. Thus, the corresponding spectra have to be identical if no spontaneous symmetry breaking occurs. Especially, there are two degenerate vector triplets.

If the discrete symmetry is spontaneously broken, then the physical spectrum is the one of standard perturbation theory, where only one triplet of vector bosons appears.

\section{Lattice simulations}\label{sec:lattice}

The previous results have been obtained under the assumption of the validity of the FMS mechanism. The prerequisite for this is that the expansion parameter is sufficiently small in average~\cite{Frohlich:1980gj,Frohlich:1981yi}. Already in the case with a single doublet this condition does not hold true for large regions of the phase diagram~\cite{Maas:2013aia,Maas:2014pba}. Especially the regions with very light and very heavy Higgs particles appears still somewhat involved.

Since the relevant parameter range for the 2HDMs is much larger without further experimental constraints, it appears therefore important to check the validity of the FMS mechanism. Lattice simulations are a possible tool, and 2HDM are accessible in such simulations~\cite{Lewis:2010ps,Maas:2014nya}. Calculating the spectrum and testing the FMS mechanism is a straightforward extension of~\cite{Wurtz:2013ova,Maas:2013aia,Maas:2014pba}, and should not pose a conceptual problem, though especially correlators of scalars are numerically expensive.

The basic approach would essentially be to simulate the 2HDM for various sets of parameters, and investigate the spectroscopy of the states listed in table \ref{table:states}, possibly supplemented by further states like in~\cite{Maas:2014pba}, as the FMS mechanism also makes statements on these. This is left to future investigations. Of course, if the FMS mechanism is not found to be working, there is no a-priori reason to expect a relation between the gauge-dependent states and the physical ones. In such a case non-perturbative calculations would anyhow be mandatory.

From a field-theoretical point of view, it would be especially interesting to investigate the cases where spontaneous symmetry breaking of the global symmetry group was assumed. In a finite lattice there is no spontaneous symmetry breaking: what we must do is to estimate the results for the infinite-volume limit and then extrapolate these estimates to the limit where there is no explicit symmetry breaking and check if this extrapolation indicates spontaneous symmetry breaking~\cite{Lewis:2010ps}. Consider the Higgs potential of the previous section and the correlation functions:\\

\noindent $<H_1^\dagger(y)H_1(y)H_1^\dagger(x)H_1(x)>$ and $<H_2^\dagger(y)H_2(y)H_2^\dagger(x)H_2(x)>$\\

\noindent After gauge fixing, we can expand them as:\\

\noindent $<H_1^\dagger(y)H_1(y)H_1^\dagger(x)H_1(x)>\approx \frac{v^4}{4}+\frac{v^2}{2}<\varphi^\dagger(y)\phi_0\phi_0^\dagger\varphi(x)>+...$\\
$<H_2^\dagger(y)H_2(y)H_2^\dagger(x)H_2(x)>=<\varphi^\dagger_2(y)\varphi_2(y)\varphi^\dagger_2(x)\varphi_2(x)>$\\

\noindent where $\varphi_2\equiv \phi_0^\dagger\Sigma_4 \varphi$. Neglecting interactions, we expect the energy spectrum of the first correlation function to start at the mass $m_h$
while for the second correlation function it should start around the mass $2m_H$. The interactions should change both correlation functions 
but not to the point where the two correlation functions are exactly equal.
This is what we intend to simulate in the limit where $\mu_5\to 0$ but always with $\mu_5> 0$ and the $Z_2$ symmetry is recovered. The same applies to the remaining propagators (and its $Z_2$ correspondents).
If from the start $\mu_5=0$, then by definition of the partition function the correlation functions are $Z_2$ symmetric.

\section{Introducing Photons}\label{sec:photons}

In this and the next section we add photons and fermion to the 2HDM, respectively, under the assumption that the FMS mechanism can be applied. In the case with one doublet, these additions can be found already in~\cite{Frohlich:1980gj,Frohlich:1981yi}.

We now consider a Lagrangian invariant under the $U(1)_Y$ gauge symmetry with generator $\Sigma_1\Sigma_2$: 
\begin{align*}
\mathcal{L}\equiv ((D^\mu+\Sigma_1\Sigma_2\frac{g'}{2}B^\mu)\phi)^\dagger(D_\mu+\Sigma_1\Sigma_2\frac{g'}{2}B_\mu\phi)-V(\phi)-\frac{1}{4}W_{\mu\nu}^aW^{a\mu\nu}-\frac{1}{4}B_{\mu\nu}B^{\mu\nu},
\end{align*}
where the $B_\mu$ is the $U(1)_Y$ gauge field,  $B_{\mu\nu}\equiv\partial_\mu B_\nu-\partial_\nu B_\mu$ is the gauge field strength tensor and finally $g'$ is the $U(1)_Y$ coupling constant\footnote{$D_\mu\equiv\partial_\mu+igW_\mu^j\frac{\sigma^j}{2}$ is defined as in the previous sections.}.

Then the background symmetry is the semi-direct product $(U(1)_Y\times Spin(3))\rtimes Z_2$ of the custodial $Spin(3)$ group whose generators are $\Sigma_3\Sigma_4$, $\Sigma_3\Sigma_5$, $\Sigma_4\Sigma_5$ (the only ones that commute with $\Sigma_1\Sigma_2$) and the $Z_2$ group generated by $\phi\to \Sigma_1\phi$. The $U(1)_Y\times Spin(3)$ is a normal subgroup.  Any transformation may be written as the product of an element of $U(1)_Y\times Spin(3)$ and an element of $Z_2$, for instance the (standard) charge reversal transformation is $\phi\to \Sigma_2\Sigma_3\phi$ and $B_\mu\to -B_\mu$.
Note that parity and charge reversal are conserved separately in the absence of fermions.

To establish contact to the usual phenomenology, we require that the vacuum is uncharged under the electromagnetic subgroup. The neutral vacuum condition is that the field configuration minimizing the potential $v\phi_0$ must be aligned along a linear combination of $\Sigma_{3,4,5}$ which all commute with the $U(1)_Y$ generator $\Sigma_1\Sigma_2$. By reparametrization we choose $\Sigma_5\phi_0=\phi_0$. We define
\begin{align*}
 H_1&\equiv\frac{1-i\Sigma_1\Sigma_2}{2}\frac{1+\Sigma_5}{2}\phi\\
H_2&\equiv \Sigma_4\Sigma_5\frac{1-i\Sigma_1\Sigma_2}{2}\frac{1-\Sigma_5}{2}\phi
\end{align*}

\noindent The previous gauge-invariant operators can be rewritten, making the mixing between hypercharge and weak isospin manifest, as
\begin{itemize}
\item $\mathcal{W}^+_\mu\equiv H^\dagger_1 i D_\mu  \Sigma_1\Sigma_3 H_1$;
\item $\mathcal{Z}_\mu\equiv\cos\theta_W H^\dagger_1 i D_\mu  H_1-\sin\theta_W \frac{g v^2}{4}  B_\mu$;
\item $\mathcal{A}_\mu\equiv\sin\theta_W H^\dagger_1 i D_\mu H_1+\cos\theta_W\frac{g v^2}{4} B_\mu$;
\item $H^\dagger_1H_1$;
\item $H^\dagger_1\Sigma_a H_2$ $(a=3,4)$;
\item $\mathcal{H}^+\equiv H^\dagger_1\Sigma_1 H_2$;
\end{itemize}
$\theta_W$ is the weak angle with $\cos\theta_W\equiv \frac{g}{\sqrt{g^2+g^{'2}}}$.

Under a gauge transformation $U(1)_Y$ where $\phi\to e^{\Sigma_1\Sigma_2 \frac{\vartheta}{2}}\phi$, we get:
\begin{align*}
\mathcal{W}^+_\mu&\to e^{i\vartheta}\mathcal{W}^+_\mu\\
\mathcal{A}_\mu&\to \mathcal{A}_\mu-\frac{1}{g\sin\theta_W}\partial_\mu\vartheta\\
\mathcal{H}^+&\to e^{i\vartheta}\mathcal{H}^+
\end{align*}
The remaining states are invariant under $U(1)_Y$. Note however that, because $U(1)_Y$ is an Abelian gauge symmetry, it is possible to provide a gauge-invariant dressing for the hypercharge (or electromagnetic) subgroup in the same way as for the standard model~\cite{Haag:1992hx}. Since this is an overall phase factor, this does not interfere with the present construction, and we therefore do not explicitly include it here.

Under charge conjugation the states transform as
\begin{align*}
\mathcal{W}^+_\mu&\to (\mathcal{W}^{+}_\mu)^*\\
\mathcal{Z}_\mu&\to -\mathcal{Z}_{\mu}\\
\mathcal{A}_\mu&\to -\mathcal{A}_{\mu}\\
\mathcal{H}^+&\to (\mathcal{H}^+)^*\\
H^\dagger_1\Sigma_3 H_2 &\to -H^\dagger_1\Sigma_3 H_2
\end{align*}
We now choose the minimum $\frac{v}{\sqrt{2}}\phi_0$ to be constant and to satisfy
\begin{align*}
\epsilon_{jkl} A_{k}A_{l}\phi_0=\epsilon_{jkl} \Sigma_{k}\Sigma_{l}\phi_0\ (j=1,2,3)
\end{align*} 
Then the minimum conserves the electromagnetic charge with generator $(\Sigma_1\Sigma_2-A_1A_2)$, that is, $(\Sigma_1\Sigma_2-A_1A_2)\phi_0=0$.

We now only consider the expansion around the vacuum. Keeping only the first non-constant terms in the expansion we get:
\begin{align*}
\mathcal{W}^+_\mu&\approx \frac{g v^2}{8} (W_\mu^1-iW_\mu^2)\\
\mathcal{Z}_\mu&\approx \frac{gv^2}{4} (\cos\theta_W W_\mu^3-\sin\theta_W B_\mu)\\
\mathcal{A}_\mu&\approx \frac{gv^2}{4} (\sin\theta_W W_\mu^3+\cos\theta_W B_\mu)\\
H^\dagger_1 H_1&\approx \frac{v^2}{2}+v\phi_0^\dagger\varphi\\
H^\dagger_1\Sigma_a H_2&\approx \frac{v}{2}\phi_0^\dagger\Sigma_a\varphi\ (a=3,4)\\
\mathcal{H}^+&\approx \frac{v}{2}\phi_0^\dagger(\Sigma_1-i\Sigma_2)\varphi
\end{align*}
On the right-hand side we can identify the states described in perturbation theory: $W_\mu^\pm\equiv \frac{1}{\sqrt{2}}(W_\mu^1\pm iW_\mu^2)$, $Z_\mu\equiv (\cos\theta_W W_\mu^3-\sin\theta_W B_\mu)$, the photon field 
$A_\mu\equiv (\sin\theta_W W_\mu^3+\cos\theta_W B_\mu)$, the charged Higgs boson $H^\pm\equiv \frac{1}{\sqrt{2}}\phi_0^\dagger(\Sigma_1\pm i\Sigma_2)\varphi$, the \ac{CP} pseudoscalar $I\equiv \phi^\dagger_0 \Sigma_3\varphi$ and finally the scalars $R\equiv \phi^\dagger_0 \Sigma_4\varphi$ and the Higgs boson $h^0\equiv \phi^\dagger_0\varphi=\phi^\dagger_0\Sigma_5\varphi$.

The multiplet $(H^0,R,I)$ transforms as a $SO(3)$ vector under custodial transformation. Also, the vacuum direction $u\equiv(\phi_0^\dagger\Sigma_5\phi_0,\phi_0^\dagger\Sigma_4\phi_0,\phi_0^\dagger\Sigma_3\phi_0)$
will transform in the same way and defines the Higgs basis.

In general the vector Higgs mass eigenstates $(h_1,h_2,h_3)$ will result from a $SO(3)$ rotation of the Higgs basis states $(H^0,R,I)$, with angles determined by the Higgs potential. Writing $h_j=n_{ja}\phi^\dagger_0 \Sigma_a\varphi$, with $n_{ja}n_{ja}=1$, the $SO(3)$ rotation $n$ relates the Higgs basis with the basis of mass eigenstates. This works as in perturbation theory.

The states are precisely as the ones expected in perturbation theory. Thus, provided the FMS mechanism works without photons, the presence of photons should not be in conflict with the FMS mechanism.

\section{Introducing Fermions}\label{sec:fermions}

The weak interactions couple to the fermions such that parity and charge reversal are not conserved separately, but only their composition symmetry CP is. The CP symmetry is then violated by the CKM matrix.

Consider a fermionic field $Q_L$ satisfying $\Sigma_1\Sigma_2Q_L=iQ_L$, $\Sigma_5Q_L=Q_L$ and transforming under the gauge symmetry $SU(2)_L$ in the same way as $\phi$. This choice for $Q_L$ already fixes $\Sigma_5$. As a consequence the most general minimum does not yet satisfy $\Sigma_5\phi_0=\phi_0$. Introduce also fermions $d_R$, $u_R$ which are singlets under $SU(2)_L$. We set the hyper-charges of the gauge symmetry $U(1)_Y$ as $Q_L(1/6_Y)$, $d_R(-1/3_Y)$, $u_R(2/3_Y)$, i.e. for $\phi\to e^{\Sigma_1\Sigma_2 \frac{\vartheta}{2}}\phi$ then $Q_L\to e^{i \frac{\vartheta}{6}}Q_L$. Hence, these are quarks.

The most general $SU(2)_L$ gauge invariant products of $\phi$ and $Q_L$ are complex linear combinations of $\overline{Q_L}\phi$, $\overline{Q_L}i\Sigma_3\phi$, $\overline{Q_L}i\Sigma_2\phi$, $\overline{Q_L}i\Sigma_1\phi$ and their hermitian conjugates\footnote{The basis of symmetric matrices commuting with the generators of $SU(2)_L$
is $\{1,\Sigma_a\}$, of skew-symmetric matrices is $\{[\Sigma_a,\Sigma_b]\}$ with $a,b,=1,...,5$, for a total of $16$ matrices.
Due to the two projectors in $Q_L$, we must divide the total by $4$ which leaves us with 4 
linearly independent products.}. The most general gauge-invariant form for the Yukawa couplings with the quarks is then
\begin{align*}
-{\mathcal{L}}_{Y_Q} &=\overline{Q_{L}}\ \Gamma_d \phi\ d_{R}+
\overline{Q_{L}}\ \Sigma_3\Sigma_1\Gamma_u \phi\ u_{R}+ \text{h.c.} \\
\Gamma_{w}&\equiv \Gamma_{w\,0}+\Gamma_{w\,1}\Sigma_3\Sigma_4+\Gamma_{w\,2}\Sigma_4\Sigma_5
+\Gamma_{w\,3}\Sigma_5\Sigma_3)
\end{align*}
with $\Gamma_{w  a}$  self-conjugate and acting as real scalars on $\phi$, where $w=u,d$ and  $a=0,1,2,3$.

The custodial $Spin(3)$ group acts on $\phi$ and $\Gamma_{w}^\dagger$ in the same way with generators $\Sigma_3\Sigma_4$, $\Sigma_4\Sigma_5$ and $\Sigma_3\Sigma_5$, such that the product $\Gamma_w\phi$ is $Spin(3)$ invariant. We thus continue using a Majorana notation for the symmetries of the Higgs potential, which appears to be working in the following, but note the remarks in~\cite{Branco:2011iw}.

In this form, we can now finally assume, without loss of generality by reparametrization of $\Gamma_{w}$, that the minimum satisfies $\Sigma_5\phi_0=\phi_0$. In this basis we define 
$H_1\equiv\frac{1-i\Sigma_1\Sigma_2}{2}\frac{1+\Sigma_5}{2}\phi$, 
$H_2\equiv \Sigma_4\Sigma_5\frac{1-i\Sigma_1\Sigma_2}{2}\frac{1-\Sigma_5}{2}\phi$,
$\widetilde{H}_j\equiv \Sigma_3\Sigma_1 H_j^*$. The Yukawa couplings for the quarks are then rewritten as:
\begin{eqnarray*}
-\frac{v}{\sqrt 2}{\mathcal{L}}_{Y_Q} &=&\overline{Q_{L}}\ H_1 M_dd_{R}+
\overline{Q_{L}}\ H_2 N_d^0 d_R+\overline{Q_{L}}\
\widetilde{H}_{1}M_u u_{R}+\overline{Q_{L}}\ \widetilde{H}_{2} N_u^0 u_{R} + \text{h.c.},
\end{eqnarray*}
where $M_{w}\equiv \Gamma_{w 0}+i\Gamma_{w 1}$, $N_{w}^0\equiv \Gamma_{w 3}+i\Gamma_{w 4}$. The matrices $M_d\equiv U_L \diag(m_d,m_s,m_b)U_R^{d\dagger}$ and $M_u\equiv U_LV^\dagger \diag(m_u,m_c,m_t)U_R^{u\dagger}$  are the quark mass matrices and $N_{d,u}^0$ are matrices not necessarily diagonal in the quark mass eigenstate basis which may induce Higgs mediated \aclp{FCNC} at tree level. The conventional \ac{CKM} matrix is given by $V$. Note that color is not treated here explicitly, but due to confinement any observable states involving quarks or gluons are anyhow color-singlets.

There is a correspondence between the standard gauge dependent fields and the $SU(2)_L$ gauge-invariant ones, as in the 1HDM case \cite{Frohlich:1980gj,Frohlich:1981yi}. The composite operators to be considered
\begin{align*}
H_1^\dagger Q_L&\to e^{-i\frac{1}{3}\vartheta}H_1^\dagger Q_L\\
\tilde{H}_1^\dagger Q_L&\to e^{i\frac{2}{3}\vartheta}\tilde{H}_1^\dagger Q_L
\end{align*} 
still retain their gauge-dependence under the Abelian part, with the indicated transformation with $e^{i \frac{\vartheta}{2}}\in U(1)_Y$. Their electromagnetic properties are thus the same as for the elementary states.

The corresponding leading terms of the expansion after gauge fixing are proportional to: 
\begin{align*}
d_L&\equiv \phi_0^\dagger Q_L\\ 
u_L&\equiv -(\phi_0^\dagger)^*\Sigma_3\Sigma_4 Q_L
\end{align*}
Again, because both Higgs doublets are Lorentz scalars, the Lorentz quantum numbers of the composite and elementary states agree.

The lepton sector with three right handed neutrinos is analogous in the absence of Majorana masses\footnote{Promoting the $M_{u,d}$ and $N_{u,d}^0$ matrices to background fields, there is an additional background flavor symmetry for the quarks $SU(3)_Q\times SU(3)_U\times SU(3)_D$  and for the leptons in the absence of Majorana masses $SU(3)_\ell\times SU(3)_e\times SU(3)_\nu$ and also a background CP(charge-parity) symmetry. There is also an Abelian background symmetry $U(1)^3$ in addition to the global symmetry $U(1)_{nb}\times U(1)_{nl}$ related to the baryonic and leptonic (no Majorana masses) numbers.
Since the Majorana mass terms in seesaw type I  are gauge singlets, the FMS mechanism can also be extended to models with seesaw type I~\cite{nuMSM}.}, with the \ac{PMNS} matrix replacing the \ac{CKM} matrix. The argumentation for them goes hence through unchanged, and will not be repeated. 

\section{Summary}\label{sec:conclusions}

The demand of gauge invariance of physical observables must be taken directly as a demand on the spectrum of any theory. In case of the standard model, the FMS mechanism justifies that the spectrum can nonetheless be rather well described by the spectrum of the gauge-dependent elementary states. Especially, this both explains the success and justifies the use of perturbation theory in the electroweak sector. If this would not be the case, the description of the physical states would, as in QCD, require non-perturbative methods, even at weak coupling.

Here, we have investigated the two-Higgs-doublet extension of the Standard Model in the light of these insights, and extended the FMS mechanism to it. Assuming its validity, we show that under some assumptions the physical spectrum is expected to coincide with the one of the elementary states, as obtained in perturbation theory. These assumptions are that the field fluctuations around the vacuum are small in average and that there is spontaneous symmetry breaking of the global symmetry group whenever the gauge orbit minimizing the Higgs potential is not unique.

To confirm that indeed the FMS mechanism is applicable and that the assumptions are valid requires genuine non-perturbative calculations. Since a failure would have substantial impact on the phenomenological relevance of these models they are a mandatory next step.

\paragraph{Acknowledgments}
L.\ P.\ acknowledges the hospitality of the Institute of Physics at the University of Graz, where most of this work has been done, and of the
Centro de F\'isica Te\'orica de Part\'iculas at the Universidade de Lisboa.
L.P. acknowledges Gustavo Branco, Margarida Rebelo and Renato Fonseca for useful conversations.

\appendix

\section{Minima structure of the Spin(4) symmetric potential}\label{a:minima}
Defining $\Phi_1\equiv\frac{1+\Sigma_5}{2}\phi$ and $\Phi_2\equiv \Sigma_4\frac{1-\Sigma_5}{2}\phi$,
the most general $Spin(4)$ symmetric potential can be rewritten as
\begin{align*}
V(\Phi_1,\Phi_2)=&-\mu_0(\Phi_1^\dagger \Phi_1+\Phi_2^\dagger \Phi_2)-\mu_5(\Phi_1^\dagger \Phi_1-\Phi_2^\dagger \Phi_2)\\
&+\frac{1}{2}\lambda_{00}(\Phi_1^\dagger \Phi_1+\Phi_2^\dagger \Phi_2)^2+\frac{1}{2}\lambda_{55}(\Phi_1^\dagger \Phi_1-\Phi_2^\dagger \Phi_2)^2
\end{align*}
The most general gauge orbit minimizing the potential verifies $u_{a}\Sigma_a\phi=\phi$, breaking the generators of $Spin(4)$ which do not commute with $u_{a}\Sigma_a$, 
where $u$ is a vector representation of $SO(5)$ normalized to $u_au_a=1$.
Without lost of generality, we can choose a basis such that $u_{1}=u_{2}=u_{3}=0$. 
Then for $u_{4}\neq 0$ the symmetry conserved by the minimum is $Spin(3)$ with generators $\epsilon_{jkl}\Sigma_k\Sigma_l$ 
and there are three spontaneously broken generators of $Spin(4)$ namely $\Sigma_j\Sigma_4$, so according to Goldstone's theorem we expect $3$ massless 
Goldstone bosons, as was confirmed in~\cite{Lewis:2010ps}. To avoid breaking the $Spin(4)$ group spontaneously we assume from now on that  $u_{4}= 0$, i.e. 
the gauge orbit minimizing the potential verifies $\pm \Sigma_5\phi=\phi$.
 
Positivity at large field amplitudes requires $\lambda_{00}>0$ and $\lambda_{00}+\lambda_{55}>0$.
From the stability conditions $\frac{1}{2}v^2_j\equiv\frac{\mu_0+\epsilon_j\mu_5}{\lambda_{00}+\lambda_{55}}>0$, i.\ e.\ $\frac{1}{2}v_2^2=\frac{1}{2}v^2_1-2\frac{\mu_5}{\lambda_{00}+\lambda_{55}}$,
and the minima (second derivative) conditions $m_{hj}^2\equiv 2(\mu_0+\epsilon_j\mu_5)>0$ and $m_{Hj}^2\equiv 2\frac{ \epsilon_j\mu_5\lambda_{00}-\lambda_{55}\mu_0}{\lambda_{00}+\lambda_{55}}>0$.
The value of the minimum is $V(\phi^j)=-\frac{(\mu_0+\epsilon_j\mu_5)^2}{2(\lambda_{00}+\lambda_{55})}<0$.

We now look for further critical orbits for which both $u_1\equiv \Phi_1^\dagger\Phi_1>0$ and $u_2\equiv \Phi_2^\dagger\Phi_2>0$.
These will satisfy the stability conditions $-\mu_0-\mu_5+\lambda_{00}(u_1+u_2)+\lambda_{55}(u_1-u_2)=0$ and $-\mu_0+\mu_5+\lambda_{00}(u_1+u_2)-\lambda_{55}(u_1-u_2)=0$, therefore $u_1= \frac{\lambda_{55}\mu_0 + \lambda_{00}\mu_5}{2\lambda_{00}\lambda_{55}}$ and $u_2=\frac{\lambda_{55}\mu_0 - \lambda_{00}\mu_5}{2\lambda_{00}\lambda_{55}}$. The potential is $V(u_1,u_2)=-\frac{\mu_0^2}{2\lambda_{00}}-\frac{\mu_5^2}{2\lambda_{55}}$
The determinant of the Hessian matrix for the variables $(u_1,u_2)$ is $2\lambda_{00}\lambda_{55}$.

We now look for the critical orbit $\Phi_1^\dagger\Phi_1=0$ and $\Phi_2^\dagger\Phi_2=0$.
The Hessian matrix is diagonal with entries $-2(\mu_0+\mu_5)<0$ and $-2(\mu_0-\mu_5)$, which have the opposite signs of $v_1^2$ and $v_2^2$ respectively.

Without loss of generality we assume $\mu_5\geq 0$ (we can change its sign by an interchange $\Phi_1\leftrightarrow \Phi_2$).
So we have $v_2^2\leq v_1^2$ and if $v_2^2>0$ then  $V(\phi^2)\geq V(\phi^1)$, so for the first orbit $\phi^1$ there is an absolute minimum.
We identify $v\equiv v_1$, $m_h\equiv m_{h1}$, $m_H\equiv m_{H1}$ and write $\lambda_{00}=(m_h^2+m_H^2-2\mu_5)/v^2$, $\lambda_{55}=(-m_H^2+2\mu_5)/v^2$ and $\mu_0=\frac{m_h^2}{2}-\mu_5$.

The conditions $u_1=-\frac{m_{H2}^2m_h^2}{4v^2\lambda_{00}\lambda_{55}}>0$ and $u_2=-\frac{m_H^2m_h^2}{4v^2\lambda_{00}\lambda_{55}}>0$ imply $\lambda_{55}<0$ and  $m_{H2}^2>0$ so if $(u_1,u_2)$ is a stability point it 
is necessarily a saddle point since the determinant of the Hessian matrix for the variables $(u_1,u_2)$ is $2\lambda_{00}\lambda_{55}<0$.
In that case we have that $V(u_1,u_2)-V(\phi^1)=-\frac{m_H^4m_h^2}{8v^2\lambda_{00}\lambda_{55}}>0$ as expected.

We also have $\frac{v_2^2}{v^2}=(1-4\frac{\mu_5}{m_h^2})$.
As for the Hessian matrix for $\phi^2$, we have
$m_{h2}^2=m_h^2-4\mu_5$ with the same sign as $\frac{v_2^2}{v^2}$ and 
\begin{align*}
m_{H2}^2&=8\mu_5^2/m_h^2-4\mu_5/m_h^2(m_h^2+m_H^2)+m_H^2,
\end{align*}
i.\ e.\ $m_{H2}^2=\frac{1}{2 m_h^2}(4\mu_5-m_h^2-m_H^2-\sqrt{m_H^4+m_h^4})(4\mu_5-m_h^2-m_H^2+\sqrt{m_H^4+m_h^4})$ .

Note that $m_h^2-\sqrt{m_H^4+m_h^4}<0$ which implies $m_H^2+m_h^2-\sqrt{m_H^4+m_h^4}<2m_H^2<m_H^2+m_h^2+\sqrt{m_H^4+m_h^4}$.

We have the following possibilities:

\begin{itemize}
\item for $(m_H^2+m_h^2+\sqrt{m_H^4+m_h^4})<4\mu_5< 2(m_h^2+m_H^2)$ then  $m^2_{h2}<0$ and $m^2_{H2}>0$ and $u_1<0$:
two critical orbits ($(v^2/2,0)$ absolute minimum and $(0,0)$ saddle);

\item for $m_h^2<4\mu_5<(m_H^2+m_h^2+\sqrt{m_H^4+m_h^4})$ then  $m^2_{h2}<0$ and $m^2_{H2}<0$:
two critical orbits ($(v^2/2,0)$ absolute minimum and $(0,0)$ saddle);

\item for $m_H^2+m_h^2-\sqrt{m_H^4+m_h^4}<4\mu_5<m_h^2$ then  $m^2_{h2}>0$ and $m^2_{H2}<0$:
three critical orbits ($(v^2/2,0)$ absolute minimum, $(0,v_2^2/2)$ saddle and $(0,0)$ local maximum);

\item for $0<4\mu_5<m_H^2+m_h^2-\sqrt{m_H^4+m_h^4}$ then $m^2_{h2}>0$ and $m^2_{H2}>0$ and $u_1>0$:
four critical orbits ($(v^2/2,0)$ absolute minimum, $(0,v_2^2/2)$ local minimum, $(u_1,u_2)$ saddle and $(0,0)$ local maximum);
\end{itemize}

\addcontentsline{toc}{section}{References}
\footnotesize
\singlespacing 
\bibliography{Poincare}
\bibliographystyle{utphysMM}
\begin{acronym}[nuMSM]
\acro{2HDM}{two-Higgs-doublet model}
\acro{ATLAS}{A Toroidal LHC ApparatuS}
\acro{BR}{Branching Ratio}
\acro{BGL}{Branco\textendash{}Grimus\textendash{}Lavoura}
\acro{BSM}{Beyond the Standard Model}
\acro{CL}{Confidence Level}
\acro{cLFV}{charged Lepton Flavor Violation}
\acro{CLIC}{Compact Linear Collider}
\acro{CMS}{Compact Muon Solenoid}
\acro{CP}{Charge-Parity}
\acro{CPT}{Charge-Parity-Time reversal}
\acro{DM}{Darkmatter}
\acro{EDM}{Electric Dipole Moment}
\acro{EFT}{Effective Field Theory}
\acro{EW}{Electroweak}
\acro{EWSB}{Electroweak symmetry breaking}
\acro{FCNC}{Flavour Changing Neutral Current}
\acro{MET}{Missing Transverse Energy}
\acro{MFV2}{Minimal Flavor Violation with two spurions}
\acro{MFV6}{Minimal Flavor Violation with six spurions}
\acro{GIM}{Glashow\textendash{}Iliopoulos\textendash{}Maiani}
\acro{GNS}{Gelfand-Naimark-Segal}
\acro{GUT}{Grand unified theory}
\acro{ILC}{International linear collider}
\acro{LEP}{Large electron\textendash{}positron collider}
\acro{LFC}{Lepton flavor conservation}
\acro{LFV}{Lepton Flavor Violation}
\acro{LHC}{Large Hadron Collider}
\acro{MFV}{Minimal flavour violation}
\acro{MIA}{Mass insertion approximation}
\acro{MSSM}{Minimal Supersymmetry Standard Model}
\acro{nuMSM}[$\nu$MSM]{minimal extension of the Standard Model by three right-handed neutrinos}
\acro{PS}{Pati-Salam}
\acro{PT}[$\mathrm{p_T}$]{transverse momentum}
\acro{QCD}{Quantum chromodynamics}
\acro{RG}{Renormalization group}
\acro{RGE}{Renormalization group equation}
\acro{SM}{Standard Model}
\acro{SUSY}{Supersymmetry, Supersymmetric}
\acro{VEV}{Vacuum expectation value}
\acro{MEG}{Muon to electron and gamma}
\acro{NP}{New Physics}
\acro{NH}{Normal hierarchy}
\acro{IH}{Inverted hierarchy}
\acro{CKM}{Cabibbo\textendash{}Kobayashi\textendash{}Maskawa}
\acro{PMNS}{Pontecorvo-Maki-Nakagawa-Sakata}

\end{acronym}
\normalsize
\onehalfspacing 
\end{document}